\documentstyle[editedvolume,numreferences]{crckapb}
\input epsf

\begin{opening}
\title{Formation, Interaction and Observation of
Topological Defects}

\author{Tanmay Vachaspati}
\institute{Physics Department, Case Western Reserve University\\
           Cleveland, OH 44106-7079, USA.}

\end{opening}

\runningtitle{Formation, Interaction and Observation...}

\begin{document}




In these lectures, I describe the formation of 
defect distributions in first-order phase transitions,  
then briefly discuss the relevance of defect interactions 
after a phase transition and the observational signatures of cosmic 
strings. Some open questions are also discussed.

\section{Formation of Defects}

A number of talks at this school have been dedicated to the
density of topological defects formed at a phase transition.
Here, I will mostly be concerned with the distribution of defects
formed at a phase transition. 

I will start with a description of the usual procedure to study 
the formation of defects (focussing on the case of strings),
point out some shortcomings, and then move on to describe
the connection of the defect formation problem with froths
and percolation. An enormous range of problems is still left 
untouched and I will end up by describing some of these.

\subsection{$U(1)$ string network: conventional analyses}

The formation of strings can be studied numerically
by assigning the $U(1)$ phase, $\alpha$, randomly on lattice
sites - say of a cubic lattice. Then one can evaluate differences 
in $\alpha$ along the edges of each plaquette of the lattice. 
To do this, it is necessary to
interpolate between the values of $\alpha$ at two neighbouring
sites. Then one finds the integral 
\begin{equation}
\Delta \alpha \equiv \oint d\alpha
\label{alphaint}
\end{equation}
around a plaquette. If this is non-zero, it indicates
that there is a string or anti-string passing through
the plaquette. In this way, all the strings are found.
Then they are connected and information about the
distribution of string is stored.

The surprising result that emerges from numerical simulations
\cite{tvav} is that most of the energy in the string network is
in infinite strings. Furthermore, the strings are Brownian on
large scales, and the loop distribution is scale invariant. Let
us explain these results in more detail:

\begin{itemize}
\item {\it Brownian strings:}
This means that the length $l$ of a string is
related to the end-to-end distance $d$ by
\begin{equation}
l = {{d^2} \over \xi}
\label{brownian}
\end{equation}
where $\xi$ is a length scale also called the step length which would
be roughly given by the lattice spacing. This result is 
valid for large $l$ to a good approximation. (The numerical results
give an exponent of about 1.92 rather than 2.00.) 
For smaller lengths, the walk is not Brownian
as lattice effects are present.

\smallskip

\item {\it Loop distribution:} Scale invariance means that there is no
preferred length scale in the problem apart from the lattice cut-off.
Then, scale invariance would imply that the number density of 
loops having size between $R$ and $R+dR$ is given by dimensional analysis:
$$
dn(R) = c {{dR} \over {R^4}}
$$
where $c\sim 6$. Using eq. (\ref{brownian}), this
may be written as:
$$
dn(l) = {{c} \over {2\xi^{3/2}}} {{dl}\over {l^{5/2}}} \ .
$$
Note that the scale invariance is in the {\it size} of the loops
and not in their {\it length}. (This is because, if we were to
examine the network with a magnifying glass, the sizes of the 
loops would be rescaled by the magnification factor but not their 
lengths.)

\smallskip

\item {\it Infinite strings:} With the implementation of the geodesic
rule, the density in infinite strings was estimated
to be about 80\% of the total density in strings. The way this
estimate was made \cite{tvav}
was to perform the simulation on bigger and bigger
lattices and to keep track of the length in the strings that were
longer than a large critical length (compared to the lattice size). 
As the lattice was made bigger, the fraction of string in long strings
tended to stabilize around 80\%. Simulations on other lattices and with
periodic boundary conditions also yield infinite strings but the
estimated fraction can vary upward from about 74\%. Analytic estimates
of the fraction which assume that the strings are random walks on a lattice,
are consistent with these estimates \cite{scherrerfrieman}.
\end{itemize}

Can one analytically see the presence of infinite strings?
This is an open question. Some progress can be made if one assumes that
strings perform a Brownian walk \cite{scherrerfrieman}. It is known that
random walks do not close in 3 dimensions and this tells us that infinite
strings will be present. Furthermore, estimates can be obtained for
the fraction of length in infinite strings and the result is similar
(though not identical) to the one obtained in simulations.

For cosmological applications such as the formation of large-scale
structure, the existence of infinite strings can be vital. The reason
is that the small closed loops can decay by emitting gravitational
and other forms of radiation but the infinite strings are destined
to live forever because of their topological character\footnote{The
two ways in which they could decay are: a) a string meets an antistring
and annihilates, and, b) a string snaps leading to a gravitational
singularity. Neither process is expected to occur at a rate that
would be cosmologically interesting.}. So only the infinite strings (and
their off-spring loops) could live to influence late time cosmology and
also to tell us the story of the cosmological phase 
transition\footnote{If the loop density is high enough, the loops 
could reconnect and lead to the formation of 
infinite strings \cite{CopKibSte}.}.

\begin{figure}[tbp]
\centerline{\epsfxsize = 0.8\hsize \epsfbox{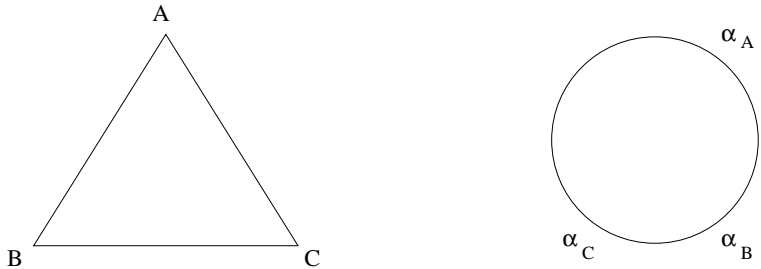}}
\caption{The triangle is a plaquette in space and the circle denotes
the vacuum manifold. At each vertex of the plaquette a phase is
assigned at random. In traversing from $A$ to $B$ on the triangle,
the phase must change from $\alpha_A$ to $\alpha_B$. However, there
is an infinite degeneracy in the path from $\alpha_A$ to $\alpha_B$
on the vacuum manifold since the path can wrap around the entire
circle any number of times.
\label{fig:algo}}
\end{figure}

A subtle issue in calculating the integral in eq. (\ref{alphaint})
is the interpolation
as we go from one lattice site to another. Consider a $U(1)$ string
simulation as shown in Fig. \ref{fig:algo}. As we traverse the triangle 
ABC in space,
the phase varies from $\alpha_A$ to $\alpha_B$ to $\alpha_C$ and then
back to $\alpha_A$. These are simply points on a circle and we know that
there are infinitely many paths joining any two points on a circle.
(The paths can go around the circle infinitely many times.) So at every
stage of the construction, we need to interpolate between the phases and
there is an infinite-fold ambiguity in this interpolation.
How do we resolve this ambiguity?

In the case of global strings, it is assumed that the shortest
of the infinitely many paths is the correct one. The rationale for
this choice is that the free energy density gets contributions from
a term $|\nabla \alpha |^2$ and this is least for the shortest
path. The rule of choosing the shortest path to interpolate
between two points on the vacuum manifold is known as the
``geodesic rule''.

In the case of gauge strings, the rationale for the geodesic
rule breaks down since the contribution to the free energy
involves the covariant derivative of $\alpha$ and not the
ordinary derivative. Now which path should be chosen?
Following the logic of the global case, it should be the
path that minimizes $|\nabla \alpha - e A|^2$, but
this would mean keeping track of the gauge field $A$ as well, 
which would make the simulation much more difficult. 

I will now discuss how one might do away with the assumptions of the
geodesic rule and the regular lattice in the simulations of string
formation.

\subsubsection{Relaxing the Geodesic Rule}

A possible cure for the ambiguity in choosing the path on the
vacuum manifold (discussed above) is to relax
the geodesic rule and
assume that the phase difference between two lattice sites is given
by a probability distribution \cite{lptv}.
If the values of the phases at lattice sites 1 and 2 are
$\alpha_1$ and $\alpha_2$, the phase difference will be
$$
\Delta \alpha = \alpha_2 - \alpha_1 + 2\pi n
\equiv \delta \alpha + 2\pi n
$$
where $n$ is a random integer drawn from some distribution. A
convenient choice for the distribution is
\begin{equation}
P_n = \int_{n-0.5}^{n+0.5} dm ~
2 \sqrt{\pi \beta} e^{-\beta (\delta \alpha + 2\pi m)^2} \ .
\label{pdist}
\end{equation}
with $\beta \ge 0$ being a parameter.
This probability distribution is consistent with the idea
that longer paths on the circle should be suppressed but the
amount of suppression depends on the choice of $\beta$. Note
that $\beta$ plays the role of inverse temperature since
lower values of $\beta$ (that is, higher temperatures) allow
for larger values of $n$ while larger values of $\beta$ reduce
the algorithm to the geodesic rule.

\begin{figure}[tbp]
\centerline{
\epsfxsize = 2.25 in \epsfbox{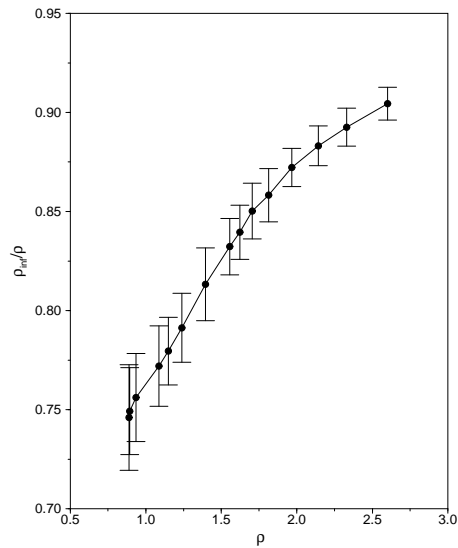}
}
\vskip 1.5 cm
\caption{A plot of the infinite string density fraction versus the total
string density. The total string density increases as the parameter
$\beta$ is lowered. The geodesic rule is recovered
in the limit that $\beta$ becomes very large.
\label{fig:relax}}
\end{figure}

In simulations that relax the geodesic rule \cite{lptv}, it is found
that the fraction of infinite strings gets larger with smaller values
of the parameter $\beta$ (see Fig. \ref{fig:relax}). 
To understand this result, note that the
smaller the value of $\beta$, the higher is the total amount of
string per plaquette because the chance of going around the 
vacuum manifold increases. Now the higher the string density, 
the more difficult is it for a string to close since there are
more ways for it to connect with other strings and wander away. 
Hence, the smaller the value of $\beta$, the higher is the fraction 
of infinite strings. Stated differently, relaxing the geodesic
rule leads to a greater fraction of infinite string.

\subsubsection{Problems with Lattice Based Simulations}

The simplest way to see that lattice based simulations might be
suspect is to realize that the critical percolation probability
depends on the lattice that is used.

Consider the case of domain walls in which one throws down one of 
two phases (+ and $-$) on the sites of a lattice. Let us denote the
probability of laying down a + by $p$. When the critical percolation
probability, $p_c$, is less than 0.5, there are three possible
phases:\\
$\bullet$ $p < p_c$: the + domains are islands in a sea of $-$.\\
$\bullet$ $p_c <p< 1-p_c$: the + and $-$ both form seas.\\
$\bullet$ $1-p_c < p$: the $-$ domains form islands in a sea of +.\\
In the unbiased case, $p=0.5$, and we get
seas of + and $-$. Then the boundary between the + and $-$ regions are
also infinite. That is, the domain walls are infinite in size.

If $p_c > 0.5$, the picture is quite different. Now we have:\\
$\bullet$ $p < 1-p_c$: the + domains are islands in a sea of $-$.\\
$\bullet$ $1-p_c <p< p_c$: the + and $-$ both form islands.\\ 
$\bullet$ $p_c < p$: the $-$ form islands in a sea of +.\\
Again, in the unbiased case, $p=0.5$ and so both the + and the
$-$ form islands. The interfaces between the islands are finite in
extent and so there are no infinite domain walls.

\begin{figure}[tbp]
\vskip 2.5 truecm
\epsfxsize = 1.25 in
\centerline{
\epsfbox{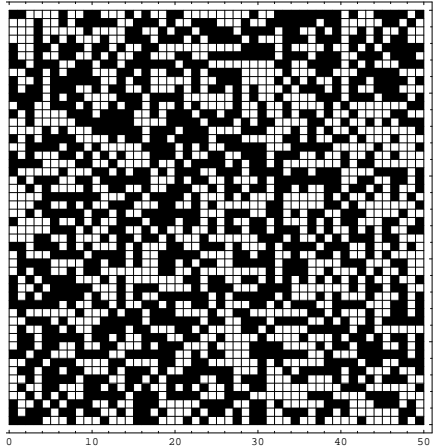}
}
\caption{The black squares denote + domains and the white squares
denote $-$ domains in a simulation with $p=0.5$ on a two-dimensional
square lattice. Neither the + nor the $-$ domains percolate in this
case and the domain walls, which are the boundaries between black
and white squares are all finite.
\label{fig:sqlat}}
\end{figure}

What is quite interesting is that, in two spatial dimensions,
$p_c=0.5$ for a triangular lattice and $p_c=0.59$ for a square
lattice. Hence the domain walls in two dimensions with $p=0.5$ are
(marginally) infinite on a triangular lattice and are all finite
on a square lattice (see Fig. \ref{fig:sqlat}). Which lattice is 
the correct one to use to study phase transitions?

One expects the same problems to arise in the lattice based study of
strings and monopoles. In fact, the study of domain walls is fundamental
to understanding strings and monopoles since strings, for example, may be
viewed as the intersection of two types of domain walls - one on which the
real part of a complex scalar field vanishes and the other on which the
imaginary part vanishes \cite{scherrervilenkin}. If the two types
of domain walls are all finite, the strings will also be finite.
Hence it is suitable to first understand the percolation of domain
walls.

\subsection{Lattice-Free Simulations}

First-order phase transitions proceed by the nucleation of bubbles of
the low temperature phase in a background of the high temperature
phase.  The bubbles then grow, collide, and coalesce, eventually
filling space with the low temperature phase. In a variety of
circumstances, the low temperature phase is not unique. Here we
will mainly consider the case where there are two low temperature
phases, which we call plus ($+$) and minus ($-$). 
Our goal is to determine the percolation probability, $p_c$. If $p_c$ 
is found to be less than 0.5, then a range of $p$ exists for which both 
the $+$ and $-$ phases will percolate and, in this case, infinite domain 
walls will be formed \cite{tvtrieste}.

\subsubsection{Random bubble lattice}

Let us begin by studying the structure of the random bubble lattice
that is produced during a first order phase transition and later
discuss percolation on this lattice. We write the bubble nucleation rate 
per unit volume as $\Gamma$, and we assume that the bubble
walls expand at constant speed $v$. From these quantities we can define 
a length scale $\xi$ and a time scale $\tau$ by:
\begin{equation}
\xi = \biggr ( {v \over \Gamma} \biggr )^{1/4} ~ , \ \ 
\tau = {1 \over {(v^3 \Gamma )^{1/4}} } ~ ,
\label{scales}
\end{equation}
where the exponents have been shown for bubbles in three dimensions. 
By rescaling all lengths (such as bubble radii) and all times by $\xi$
and $\tau$ respectively, the dependence of the problem on $\Gamma$ and
$v$ is eliminated. Therefore dimensionful quantities such as the number
density of bubbles of a given size can be rescaled to a universal distribution, 
and dimensionless quantities, such as the the critical percolation probability, 
will be independent of $\Gamma$ and $v$. 

The scaling argument given above relies on the absence of any other
length or time scales in the problem. Potentially such a scale is provided by
$R_0$, the size of bubbles at nucleation, and our assumption is that
$R_0 << \xi$. Also, note that we have taken all bubbles to expand at
the same velocity $v$. This is justified if the low temperature phases
within the bubbles are degenerate. If this degeneracy is lifted, different
bubbles can expand at different velocities and this may result in lattices
with varying properties. We are primarily interested in the exactly
degenerate case which is relevant to the formation of topological defects.


In \cite{deLVac}, the nucleation and growth of bubbles leading to the 
completion of the phase transition was simulated according to the scheme 
described in Ref. \cite{jbtktvav}. There are two ways to view this scheme. 
The first is
a dynamic view where, as time proceeds, the number of nucleation sites 
are chosen from a Poisson distribution, bubbles keep growing and 
colliding until they fill space. The second equivalent viewpoint is 
static and more convenient for simulations. A certain number of
spheres whose centers and radii are drawn from uniform distributions
are placed in the simulation box. This corresponds to a snapshot of the
bubble distribution. If the number of spheres that are laid down is
large, they will fill space and the snapshot would be at a time after
the phase transition has completed.

It is worth comparing the present model with currently existing models 
of froth.  The main distinguishing feature is that the bubbles continue to 
grow even after they collide. This is in sharp distinction with the models 
used in crystal growth such as the Voronoi and the Johnson-Mehl models. 
In these models, crystals nucleate randomly inside a volume, 
grow and then, once they
meet a neighboring crystal, stop growing in the direction of that
neighbor. (In the Voronoi model, all crystals are nucleated at one
instant while in the Johnson-Mehl model, they can nucleate at different
times.) This difference between the phase transition model and the
Voronoi type models is significant and the resulting lattices have
different properties.
Another model considered in the literature is called a
``Laguerre froth''. Here the snapshot of the domains corresponds to 
a horizontal slice of a mountain range in which each mountain 
is a paraboloid. The circles of intersection of the plane and the
paraboloids define the Laguerre froth \cite{rivier}. In terms of bubbles, 
this means that the bubble walls move with a velocity that is proportional 
to $\sqrt{t-t_0}$ where $t-t_0$ is the time elapsed since nucleation. 
Such a model in two dimensions was studied by numerical
methods in Ref. \cite{telley}. If the paraboloids are replaced by cones, 
the model comes closer to the present one. 

A feature of our model of the first order phase transition is that
bubbles cannot nucleate within pre-existing bubbles. This is
appropriate to the case where the phases existing within bubbles are
degenerate or nearly degenerate. However, in cases where a variety of 
non-degenerate bubbles can exist (for example, if the system has metastable
vacuua), this assumption may have to be relaxed \cite{gleiseretal}. 

\begin{figure}[tbp]
\centerline{
\epsfxsize = 2.65 in \epsfbox{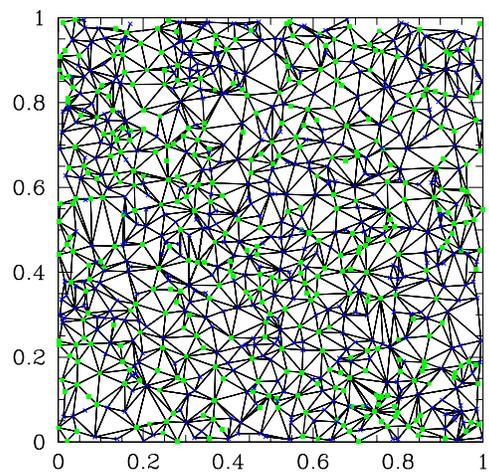}
}
\caption{The crosses denote bubble centers that are in the + phase
and the filled squares denote bubble centers that are in the - phase.
If two bubbles collide, their centers are joined by straight lines.
The figure then shows the ``random bubble lattice'' expected in a 
first order phase transition in two spatial dimensions.
\label{fig:bublat}}
\end{figure}

\begin{figure}[tbp]
\centerline{\epsfxsize = 0.8\hsize \epsfbox{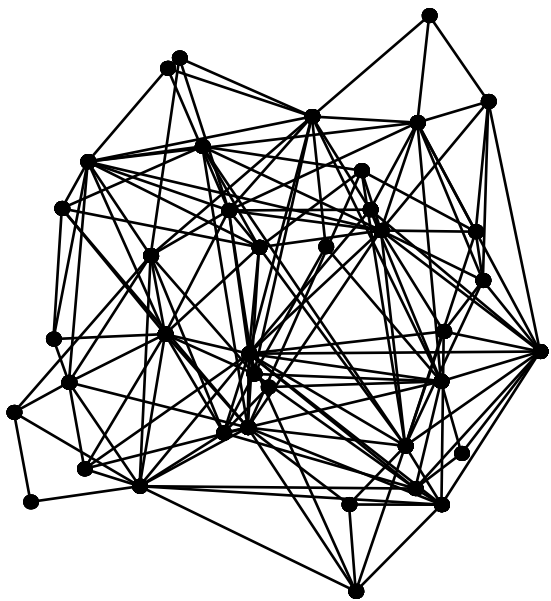}}
\caption{\label{lattice}
A portion of the three dimensional dual bubble lattice.
}
\end{figure}

A two dimensional (dual) bubble lattice is constructed by connecting
the centers of bubbles that have collided (Fig. \ref{fig:bublat}).
The three dimensional bubble lattice is similarly constructed 
and is shown in Fig. \ref{lattice}. 
The bubble lattice is almost fully triangulated though some violations of 
triangulation can occur. For example, if a tiny bubble gets surrounded by 
two large bubbles, the center of the tiny bubble will only be
connected to the centers of the two surrounding bubbles and this
can lead to plaquettes on the lattice that are not triangular.
The characteristics of this bubble
lattice hold the key to the percolation of phases and the formation of
topological defects. In particular, the average number of
vertices to which any vertex is connected is expected to play a crucial
role. This number is called the ``mean coordination number'' of the lattice,
and we now determine this quantity analytically.

First we consider the two dimensional case. We denote the number of points
in the lattice by $P$, the number of edges by $E$ and the number of
faces by $F$. Then the Euler-Poincar\'e formula \cite{nashsen} tells us
\begin{equation}
\chi = P - E + F
\label{eulerpoincare2d}
\end{equation}
where, $\chi$ is the Euler character of the lattice and is related to the
number of holes in the lattice (genus).
In our case, the lattice covers a plane which we can compactify in some
way, say by imposing periodic boundary conditions. Then $\chi$ is the
genus of the compact two dimensional surface. For us it will only be
important that $\chi = O(1)$. Next, if ${\bar z}$ is the (average) coordination 
number, we can see that 
$$
E = {{\bar z} \over 2} P ~ ,
$$
since, a given point is connected to ${\bar z}$ other points but each edge is
bounded by two points. Also,
$$
F = {2 \over 3} E ~ ,
$$
since each line separates 2 faces but then each face is bounded by 3 lines.
Now, using (\ref{eulerpoincare2d}) gives
$$
1 - {{\bar z} \over 2} + {{\bar z} \over 3} = {\chi \over P} \simeq 0 ~ ,
$$
since $P$ is assumed to be very large. Therefore, in two dimensions,
${\bar z}=6$, a result that first appeared in the botanical 
literature \cite{botany,rivier}. 

In three dimensions the analysis to evaluate ${\bar z}$ is somewhat more
complicated. The Euler-Poincar\'e formula now says
\begin{equation}
\chi = P - E + F -V ~ ,
\label{eulerpoincare3d}
\end{equation}
where $V$ is the number of volumes in the 
lattice. Now, in addition to the usual coordination number ${\bar z}$, we 
also need to define a ``mean face coordination number'' ${\bar y}$ which 
counts the average number of faces sharing a common edge. In terms of 
${\bar y}$ and ${\bar z}$, the relations between the 
various quantities for a triangulated three dimensional lattice are:
\begin{equation}
E = {{\bar z} \over 2} P ~ , \ \ 
F = {{\bar y} \over 3} E ~ , \ \ 
V = {2 \over 4} F  ~ ,
\label{reltaions}
\end{equation}
where the first equation is as in two dimensions, the second equation
follows from the definition of ${\bar y}$ and the fact that the lattice is
triangulated, and the last relation follows because a face separates
two volumes and a volume is bounded by four faces that form a tetrahedron.
Inserting these relations in (\ref{eulerpoincare3d}) leads to:
\begin{equation}
{\bar z} = {{12} \over {6-{\bar y}}}  ~ ,
\label{yzrelation}
\end{equation}
where, as before, we assume that $P$ is very large and ignore the
$\chi /P$ term.
Note that the relation between ${\bar y}$ and ${\bar z}$ is purely topological and
will hold for any triangulated lattice. 

We now want to estimate ${\bar y}$. For this we work in a ``mean
field'' approximation where we assume that the edge lengths 
are fixed. We consider two vertices $A$ and $B$ separated
by a unit distance. We wish to find the number of points that can
be connected to both $A$ and $B$, subject to the constraint that the
connected points are at unit distance from each other. This will give
the (average) number of faces that share the edge from $A$ to $B$ and
hence will be the face coordination number ${\bar y}$.  Let us choose
$A$ to be at the center of a sphere of unit radius and $B$ to be at
the North pole. Then the additional points $P_1$,...,$P_y$, have to
lie on the circle at latitude 60 degrees to satisfy the distance
constraint. Then one finds that the azimuthal angular separation of
two sequential points $P_i$ and $P_{i+1}$ is 70.5 degrees. Therefore
\begin{equation}
{\bar y} = {{360} \over {70.5}} = 5.1  ,
\label{yestimate}
\end{equation}
which then leads to \cite{coxeter}
\begin{equation}
{\bar z} = 13.4 ~ .
\label{zestimate}
\end{equation}

It is worth noting the ingredients that have entered into the analytic
estimate of ${\bar z}$. The relation (\ref{yzrelation}) is a
topological statement about the lattice, but the estimate for ${\bar
y}$ is geometric, depending on the assumption that the 
edges have fixed length. In principle, the edge lengths can 
fluctuate but our estimate for ${\bar y}$ will still 
be valid if the fluctuations average out. 

\begin{figure}[tbp]
\centerline{\epsfxsize = 0.6 \hsize \epsfbox{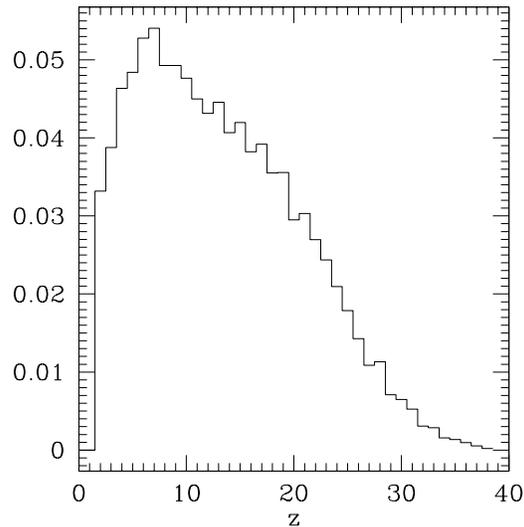}}
\caption{\label{neighbor}
The coordination number relative frequency 
distribution for the three dimensional dual bubble lattice.
}
\end{figure}

In Fig. \ref{neighbor} we show the distribution of coordination number
in our three dimensional simulations. The average coordination number
is found to be ${\bar z}=13.34 \pm 0.05$ and agrees quite closely with 
the mean field result. For comparison, Voronoi foam has ${\bar z} = 15.54$
and the Johnson-Mehl model has ${\bar z} > 13.28$ \cite{meijring}. 
The reason why ${\bar z}$ is larger in the Voronoi model is that, 
in this model, the cells stop growing on collision in the direction of
the collision, thus leading to anisotropic growth. It can be shown
that anisotropy of the cells leads to a higher value of ${\bar z}$ 
\cite{rivier}.

The mean value of $z$ is not a good characteristic of the distribution
of $z$ since the distribution is skewed (the modal value of $z$ is 7)
and it is of interest to characterize the entire distribution of $z$. 
In the literature on domain physics, attempts to derive the distribution 
of coordination number are often based on maximizing the ``entropy'' 
of the lattice. The expression used for the entropy is the one proposed 
by Shannon \cite{shannon,rivier2}. 
\begin{equation}
S = - \sum_n p_n \ln p_n
\label{entropy}
\end{equation}
where, $p_n$ is the probability that a vertex will be connected to $n$ 
other vertices. In addition, one needs to insert the Euler-Poincar\'e 
constraint in the Shannon entropy via Lagrange multipliers. (There
is also the issue of assigning {\it a priori} probabilities \cite{rivier2}.)
In two dimensions the constrained extremization is relatively 
straightforward, since the Euler-Poincar\'e constraint fixes the
average coordination to be 6, {\it i.e. }
$$
\sum_n np_n = 6 \ \ ({\rm in \ 2D}) \ .
$$

In three dimensions, we may once again introduce the constraint 
that the average coordination number of the lattice is fixed (even
though there is some freedom in choosing its value).
Then, on extremizing $S$ in eq. (\ref{entropy}) with respect to $p_n$, 
we find an exponential fall-off of the distribution. Indeed, the 
distribution shown in 
Fig. \ref{neighbor} has an exponential fall-off:
\begin{equation}
f(z) \sim \exp [ -0.25 z ] \ , \ \ z  > 20 \ .
\label{falloff}
\end{equation}

\subsubsection{Percolation on a random bubble lattice}

\begin{figure}[tbp]
\centerline{\epsfxsize = 0.6\hsize \epsfbox{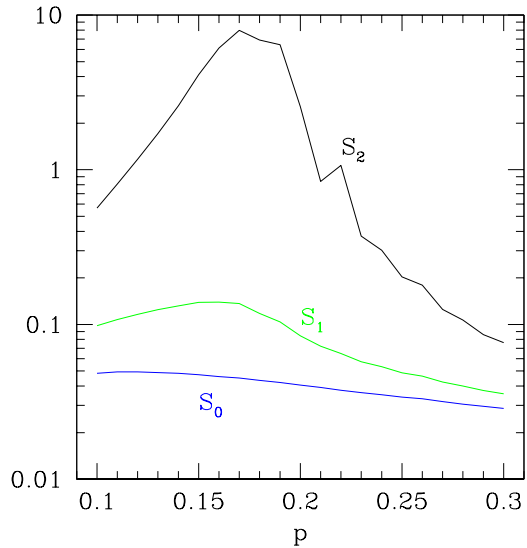}}
\caption{\label{perc}
The zeroth, first and second moments of the cluster distribution function
versus the probability $p$ in three dimensions. 
}
\end{figure}

We now turn to the formation of defects on the bubble lattice. We 
put a + phase on a bubble with probability $p$ and a $-$ phase with 
probability $1-p$ (as shown in Fig. \ref{fig:bublat} in the
two dimensional case). We then find the size distribution of + clusters 
and calculate the moments of the cluster distribution function after 
removing the largest cluster from the distribution \cite{stauffer}. 
That is, we calculate:
\begin{equation}
S_l (p) = \sum_{s \ne s_{max}} s^l  n_s (p)
\label{momentsdefn}
\end{equation}
for $l=0,1,2,...$, where the sum is over cluster sizes ($s$) but does
not include the largest cluster size, and $n_s (p)$ is the number of
clusters of size $s$ divided by the total number of bubbles.  In
Fig. \ref{perc} we show the first three moments as a function of $p$,
where the turning point in $S_2$ marks the onset of percolation. To
understand this, first consider the behavior of the second moment for
small $p$. As we increase $p$, there are fewer + clusters (as seen
from the $S_0$ graph) probably due to mergers, but the merged cluster
sizes are bigger (as seen from the $S_1$ graph).  Since the second
moment places greater weight on the size of the cluster than on the
number density as compared to the lower moments, it grows for small $p$. 
For large $p$, however, as we
increase $p$ further, the additional + clusters join the largest
cluster of +'s and are not counted in the second moment. In fact, some
of the smaller clusters also merge with the largest cluster and get
removed from the sum in (\ref{momentsdefn}). This causes the second
moment to decrease at large $p$. Hence, the second moment has a
turning point and the location of this turning point at $p_c$ marks
the onset of percolation. In three dimensions we find $p_c =
0.17\pm 0.01$ (from Fig. \ref{perc}), which is well under 0.5, 
while in two dimensions we find $p_c =0.50 \pm 0.01$ which is
consistent with 0.50. (The two dimensional version of Fig. \ref{perc}
may be found in \cite{tvtrieste}.)

It is interesting to compare the critical probabilities we have found
with lattice based results for site percolation where the regular lattice 
has a coordination number close to that of the random bubble lattice. In
two dimensions a triangular lattice has ${\bar z}=6$ and $p_c
=0.5$. In three dimensions, a face centered cubic lattice has ${\bar
z}=12$ and $p_c = 0.198$ \cite{stauffer}.  These values of the
critical probabilities are fairly close to our numerical results.

\subsubsection{Open Problems}

\noindent{\it Exponents:}
The value of the critical percolation probability is not universal. 
However, the percolation exponents are expected to be universal.
These have not been evaluated for the random lattice and would be 
worth determining.

\smallskip
 
\noindent{\it Bias:}
The rather low value of $p_c$ in three dimensions means that domain
walls formed between degenerate vacua ($p=0.5$) will percolate and
almost all of the wall energy will be in one infinite
wall. Furthermore, even if the vacua are not degenerate, {\it i.e.}
there is bias in the system, infinite domain walls can still be produced.
If the properties of the bubble lattice are insensitive to small biases,
infinite domain walls would be produced for $p > 0.17$. However,
it is likely that the bubble lattice will depend on the bias in three
ways. First, the nucleation rate of bubbles of the metastable vacuum
will be suppressed compared to that of the true vacuum. Secondly, the
velocity with which bubbles of the two phases grow can be different. Thirdly,
bubbles of the true vacuum may nucleate within the metastable vacuum.
In addition to these factors, bubbles may not retain their spherical
shape while expanding due to instabilities in their growth.
The effect of these factors on the percolation probability will be
model dependent. For example, the bubble velocities will depend on the
ambient plasma, and the nucleation rates on the action of the instantons
between the different vacuua. The effect of including these factors on
the random bubble lattice and percolation has not yet been studied. 

\smallskip

\noindent{\it Other defects:}
The formation of topological strings on the random bubble lattice 
follows the algorithm described in Ref. \cite{tvav}. It is found
that about 85\% of the strings in the simulation are infinite. 
This number should be compared with earlier
static simulations of string formation which yield a slightly lower
fraction (about $80\%$). The distribution of other types of defects
should be determined.

\smallskip

\noindent{\it Phase equilibration:}
The analysis described here neglects phase equilibration processes when
domains of different phases collide. This may be justified if the 
time scale $\tau$ in eq. (\ref{scales}) 
is short compared to the typical time required for 
phase equilibration. In the case of domain walls, phase equilibration
in two colliding bubbles can only occur by the motion of the
phase separating wall across the volume of one of the bubbles. In
this case, the neglect of phase equilibration is justified if the domain
wall velocity is much smaller than the bubble wall velocity.
It would be interesting to see how the results change if phase
equilibration is important.

\section{Interactions of Defects}

Once a string network forms in any system, the strings start moving
under their tension. Inevitably string collisions occur. What happens
then?

\begin{figure}[tbp]
\centerline{\epsfxsize = 0.8\hsize \epsfbox{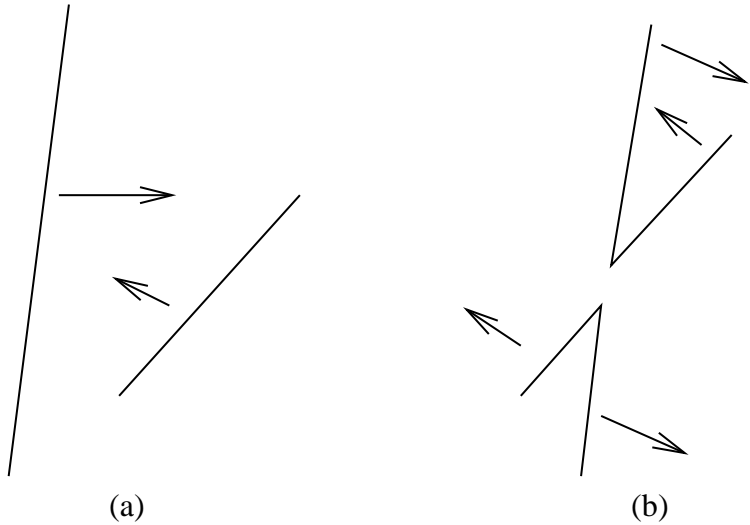}}
\caption{ \label{reconnect} Two incoming strings in (a) collide,
reconnect and emerge as in (b).
}
\end{figure}

It is a classic result that two strings intercommute (reconnect) upon 
intersection (Fig. \ref{reconnect}).  
This conclusion seems to hold regardless of the details 
of the collision (angles and velocities), as well as the physical model 
(global or local strings)\footnote{If there are strong attractive
forces between the strings and they are nearly parallel upon collision, 
they may form a bound state which smears out the collision region
and the strings may then separate again, thus passing through
each other without intercommuting \cite{exception}.}. 
Intercommuting has been observed in computer simulations
and experimentally.  There are important supporting arguments
that provide insight into intercommuting 
but there is no analytical proof that intercommuting of Abelian strings
must necessarily occur\footnote{Certain non-Abelian strings do not
intercommute but this is for topological reasons. Instead of 
intercommuting, such strings can get connected by another type
of string upon collision.}. 

The phenomenon of intercommuting is vital to the relaxation 
(coarsening) of the system. In cosmology, intercommuting provides
a means for the string network to dissipate and prevent it from
dominating the universe.

A question that has recently attracted some attention is the
interaction of {\it different types} of defects arising in the same
physical model \cite{DvaLiuVac}. Consider, for example, a phase
transition in which both domain walls and point defects (global or
local monopoles) can be formed. On formation, the domain walls will
start moving under their own tension. Inevitably, they will collide
with the point defects. What happens then? Do the point defects pass
through the walls to the other side? Or do they undergo some 
microphysical transformation? The answer to these questions are 
very important for understanding the coarsening of the system.

Based on several different arguments, it was suggested in \cite{DvaLiuVac}
that the monopoles do not pass through the domain walls. Instead
they undergo a microphysical transformation and unwind on the walls.
The arguments in support of this conjecture are:

(i) There is an attractive force between the monopoles
and the walls since monopoles can save the expense of having to
go off the vacuum in their core by moving on to the wall.
So the monopoles can form bound states with
the walls. Then, as there is no topological obstruction to the
unwinding of monopoles on the wall, the monopoles on the wall
can continuously relax into the vacuum state.

(ii) The investigation of a similar system - Skyrmions
and walls - has been dealt with in full detail in Ref. \cite{piette}.
These authors find that the Skyrmion hits the wall, sets up
traveling waves on the wall and dissipates.
They also find that, even though it is topologically possible
for the Skyrmion to penetrate and pass through the
domain wall, this never happens. They attribute their finding to the
coherence required for producing a Skyrmion. That is, the penetration
of a Skyrmion may be viewed as the annihilation of the incoming
Skyrmion on the wall and the subsequent creation of a Skyrmion on
the other side. However, the annihilation results in traveling waves
along the wall that carry off a bit of the coherence required to
produce a Skyrmion on the other side. Hence, even though there is
enough energy in the vicinity of the collision, a Skyrmion is
unable to be created on the other side of the wall. I think that
these considerations apply equally well to monopole-wall interactions
and that monopoles will never pass through the wall - just as
strings never pass through each other.

(iii) The interactions of vortices and domain walls separating
the A and B phases of $^3$He have been studied and also observed
experimentally. It is found that singular vortices do not penetrate
from the B phase into the A phase \cite{trebin}.

What is sorely needed is a direct check of this conjecture. 
(A very recent study \cite{alexander} has confirmed monopole 
dissolution on walls in a particular model under some restricted 
conditions.) If confirmed, it would be 
good to be able to understand the result at a deeper level.

\begin{figure}[tbp]
\centerline{
\epsfxsize = 0.49\hsize \epsfbox{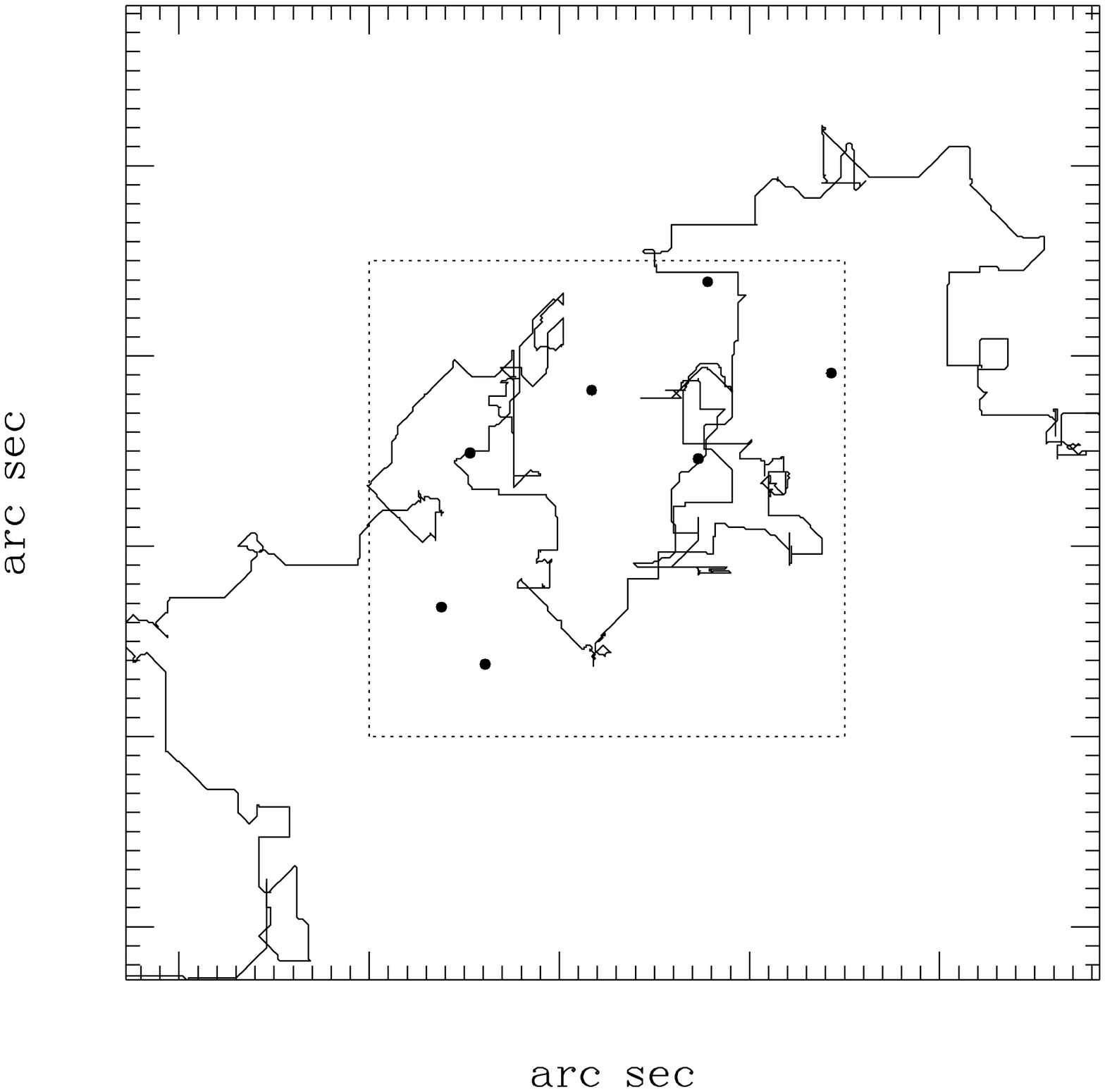}
\epsfxsize = 0.49\hsize \epsfbox{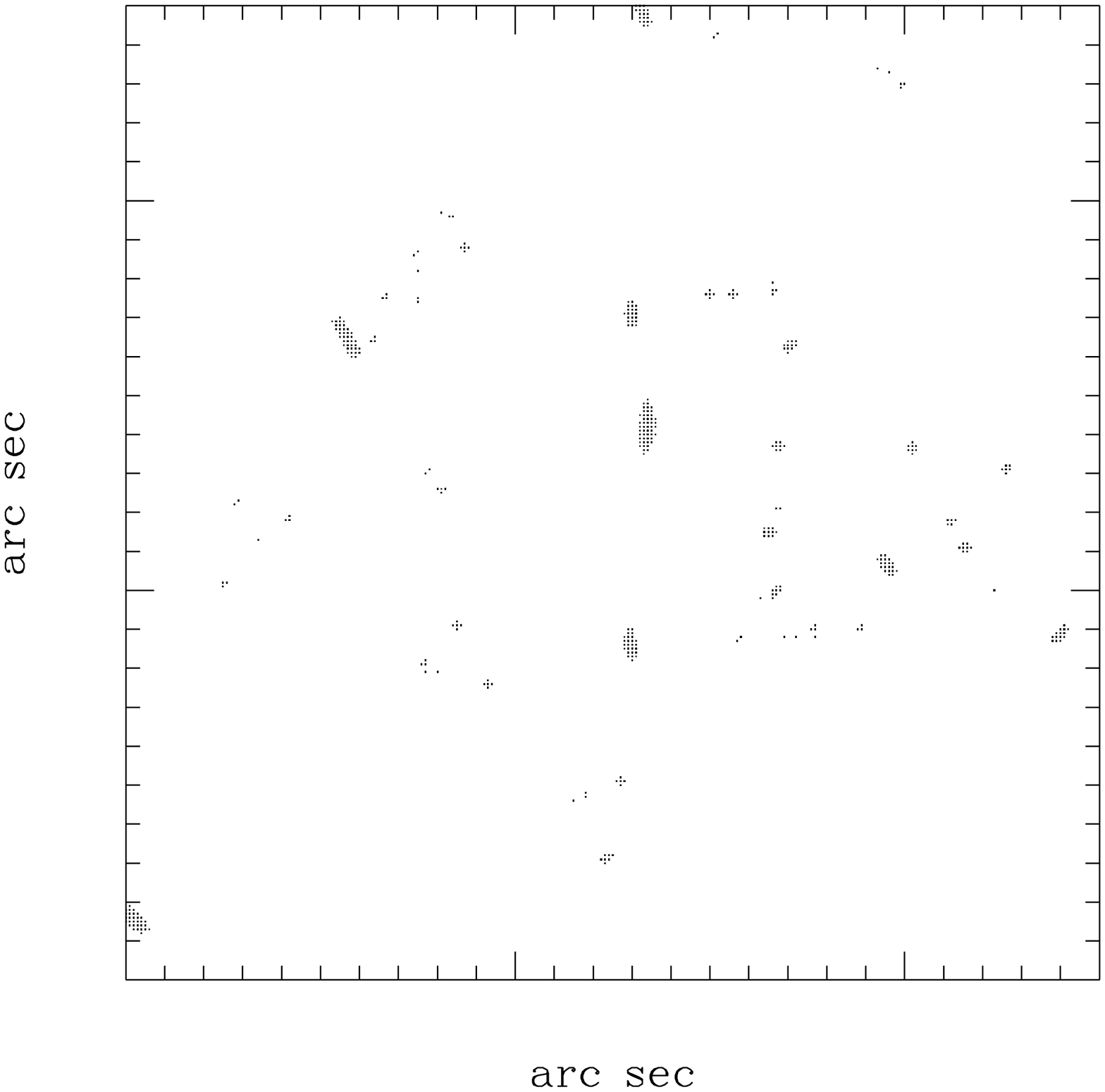}
}
\caption{\label{crit2} The figure on the left shows the
projected string configuration along with the randomly located unlensed
sources inside the dashed box.
The figure on the right shows the resulting images of the sources
The full string  shown in the figure on the left was used to determine
the lensing effects.
}
\end{figure}

\section{Observation of Cosmic Strings}

If cosmic strings lie between us and some distant sources of
light, they will cause distortions in the image of the source
or cause multiple images to be produced.
Two of the most distant cosmological sources are the cosmic microwave
background radiation and quasars. The cosmic microwave background
radiation comes to us from when the universe was only about
a million years old, while the light from quasars started out about 
1 billion years ago. (The present age of the universe is about 10 
billion years.) A major difference between these two sources is that 
the CMBR is an almost uniform source, while the quasars are point-like.
This leads to different kinds of observable signatures of strings in
the two cases. Cosmic strings can imprint a pattern of anisotropy
of the CMBR and they
can gravitationally lens quasars and galaxies with a frequency given 
by the probability that a quasar will lie behind a string. Both effects
are quite small - the anisotropy is at the level of about 1 part in $10^5$
and the probability of lensing is also of this order of magnitude.
In either case, however, the signature of strings appears to
be quite distinct and can lead to confirmation of their presence
or absence at a certain energy scale.

\subsection{Gravitational lensing}

The most tricky part in studying the gravitational lensing of distant
sources by strings is the actual construction of the string itself.
In Ref. \cite{deLKraVac}, the string was constructed by using flat
spacetime simulations as proposed by Smith and Vilenkin \cite{SmiVil}.
Light from distant sources is then propagated in the gravitational 
field of this string. The results are shown in Fig. \ref{crit2}.

What seems most striking about the lensed sources is that they
seem to trace out the string. This is due to the wiggly nature
of the string since small sections of the projected string can act
like very massive objects.

\begin{figure}[tbp]
\centerline{
\epsfxsize = 0.49\hsize \epsfbox{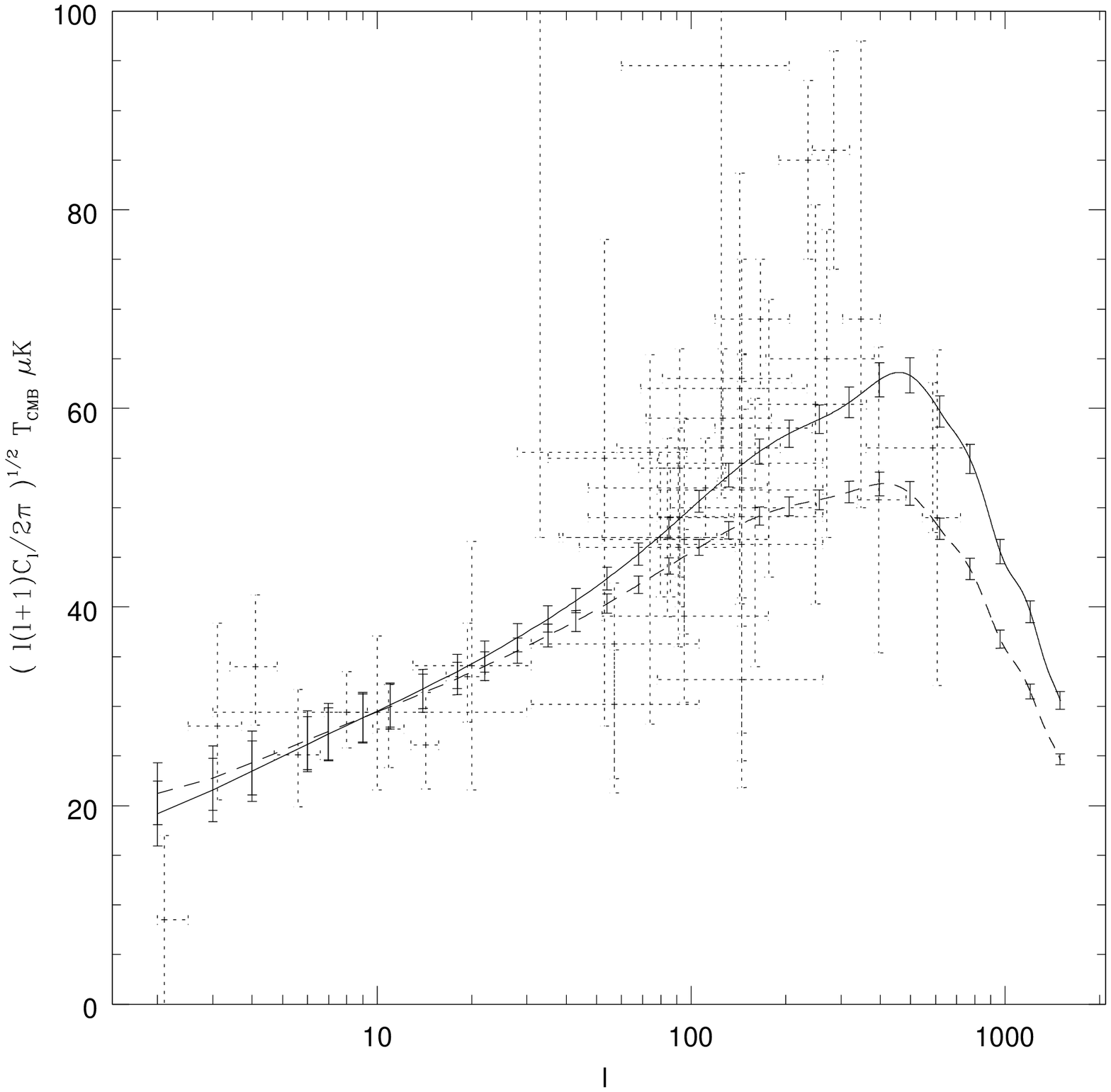}
\epsfxsize = 0.49\hsize \epsfbox{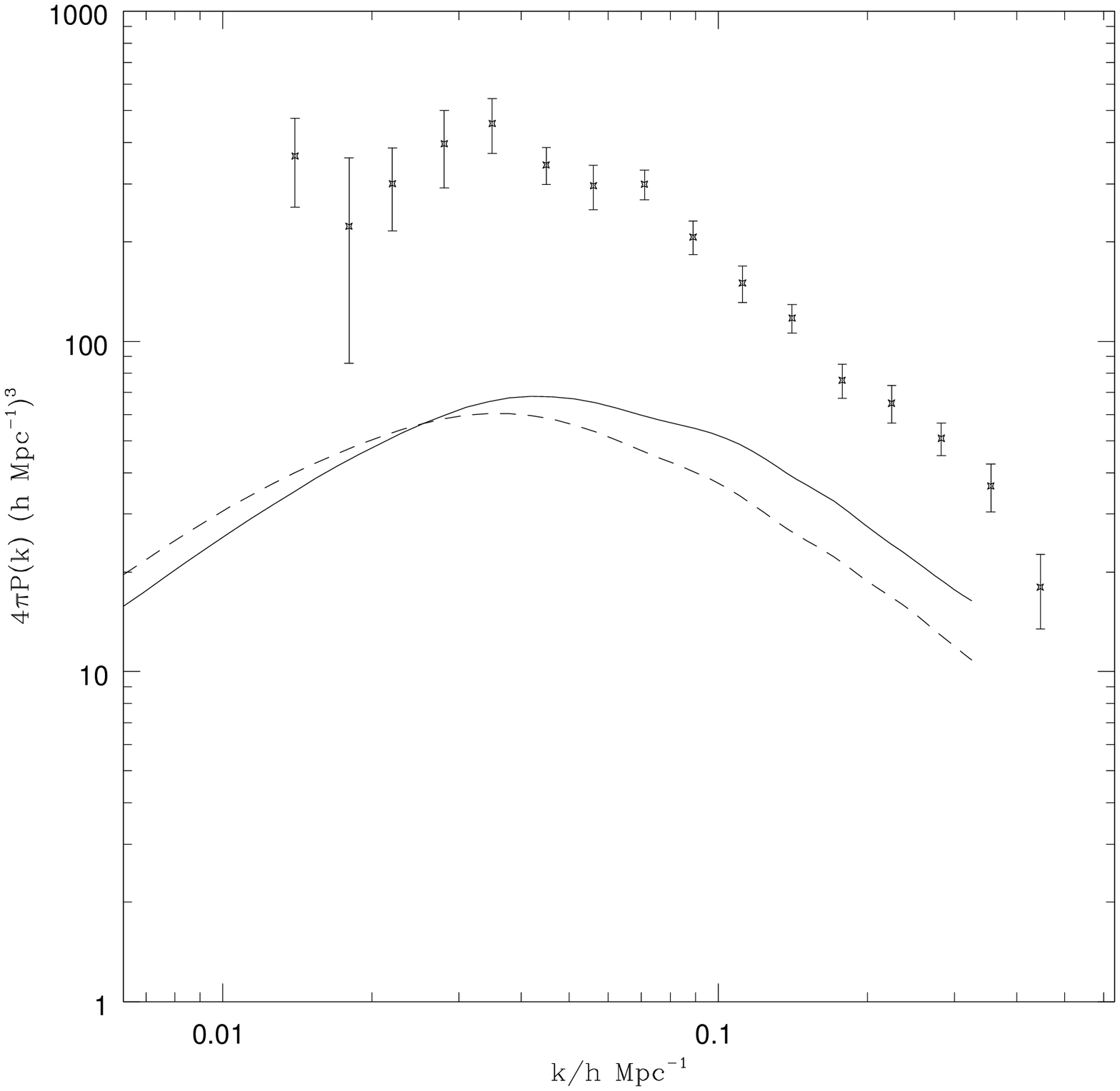}
}
\caption{
The graph on the left shows the total angular power spectrum in
CMBR anisotropy for wiggly (solid line) and smooth
(dashed line) strings when $\Omega_{baryons}=.05$, $\Omega_{CDM}=0.25$ and
$\Omega_{\Lambda}=0.7$ and using small values for the string velocities.
The graph on the right shows the power spectrum of the density
inhomogeneities for the same two models and the observed data
points.
}
\label{fig-c75}
\end{figure}

\subsection{CMBR distortions}

Once again, the most difficult aspect of calculating the CMBR
distortions is to find a reliable model of the network of strings.
Direct simulations of the string network have provided invaluable
information about certain properties of the network. The game
is to convert this information into a model that can then be
fed into the machinery to calculate the CMBR anisotropy.
Recently I have been working with Levon Pogosian to compute
the anisotropy using the model developed in \cite{Hin,Albetal}.
(Some other references to related literature may be found
in \cite{othercmbr}.)
The results depend on details of the string model as well as the
cosmological model. In Fig. \ref{fig-c75} I show 
our preliminary results for the power in the $l^{th}$ spherical 
harmonic of the CMBR anisotropy as a function of $l$ together 
with the observed data points. (Space does not permit a lengthier 
explanation of the graphs in this section but background
details can be found in several textbooks, 
for example Ref. \cite{peacock}.) At the moment, the observations
do not confirm or reject the hypothesis that cosmic strings may
be responsible for the CMBR anisotropy.

The density perturbations produced by cosmic strings would lead
to large-scale structure formation. Then one might compare the
power spectrum of the density inhomogeneities produced by
strings to those observed in the galaxy distribution. Here too,
the details of the string and the cosmological model are all
crucial. Within the limitations of the models used,
the string predictions do not agree with observations at the level
of a factor of about 2 in the amplitude of the density fluctuations 
(Fig. \ref{fig-c75}). However, further modeling of the string network
and analysis is necessary before we can be sure of this result.

\smallskip

\noindent{\bf Acknowledgments}

This work was supported by the Department of Energy (USA).



\begin{thebibliography}{999}

\bibitem{tvav} T. Vachaspati and A. Vilenkin, Phys. Rev. {\bf D30},
2036 (1984).

\bibitem{scherrerfrieman} 
R. J. Scherrer and J. Frieman, Phys. Rev. {\bf D33}, 3556 (1986).

\bibitem{CopKibSte} E. J. Copeland, T. W. B. Kibble and D. A. Steer,
Phys. Rev. {\bf D58}, 043508 (1998).

\bibitem{lptv} L. Pogosian and T. Vachaspati,
Phys. Lett. {\bf B423}, 45 (1997).

\bibitem{scherrervilenkin} 
R. J. Scherrer and A. Vilenkin, Phys. Rev. {\bf D56},
647 (1997); Phys. Rev. {\bf D58} 103501 (1998).

\bibitem{tvtrieste} T. Vachaspati, ICTP 1997 Summer School Lectures
on Cosmology, hep-ph/9710292 (1997).

\bibitem{deLVac} A. A. de Laix and T. Vachaspati, 
Phys. Rev. {\bf D59} 045017 (1999).

\bibitem{jbtktvav} J. Borrill, T. W. B. Kibble, T. Vachaspati and
A. Vilenkin, Phys. Rev. {\bf D52}, 1934 (1995).

\bibitem{rivier} See the contribution by
N. Rivier in ``Disorder and Granular Media'', eds. D. Bideau and
A. Hansen (North-Holland, 1993).

\bibitem{telley} H. Telley, Ph. D. Thesis, EPFL, Lausanne, 1989 
(unpublished).

\bibitem{gleiseretal} M. Gleiser, A. F. Heckeler and E. W. Kolb,
Phys. Lett. {\bf B405}, 121 (1997).

\bibitem{nashsen} C. Nash and S. Sen, 
``Topology and Geometry for Physicists'', Academic Press, London (1983).

\bibitem{botany} F. T. Lewis, Anat. Record {\bf 38}, 341 (1928);
{\it ibid.} {\bf 50}, 235 (1931).

\bibitem{coxeter} H. S. M. Coxeter, Ill. J. Math. {\bf 2}, 746 (1958);
J. A. Dodds, J. Coll. Interf. Sci. {\bf 77}, 317 (1980); N. Rivier,
J. Physique Coll. {\bf 43}, C9-91 (1982).

\bibitem{meijring} J. L. Meijring, Philips Res. Rep. {\bf 8}, 270 (1953).

\bibitem{shannon} C. E. Shannon, Bell Systems Technical Journal
{\bf 27}, 379 (1948); reprinted in C. E. Shannon and W. Weaver,
``The Mathematical Theory of Communication'', (University of Illinois
Press, 1949).

\bibitem{rivier2} N. Rivier, Phil. Mag. {\bf B52}, 795 (1985).

\bibitem{stauffer} D. Stauffer, Phys. Rep. {\bf 54}, 1 (1979).

\bibitem{exception} L. M. A. Bettencourt and T. W. B. Kibble,
Phys. Lett. {\bf B332}, 297 (1994). 

\bibitem{DvaLiuVac} G. Dvali, H. Liu and T. Vachaspati, 
Phys. Rev. Lett. {\bf 80}, 2281 (1998).

\bibitem{piette} A. Kudryavtsev, B. Piette, W. J. Zakrzewski,
hep-th/9709187, DTP-97/25 (1997).

\bibitem{trebin}
H.-R. Trebin and R. Kutka, J. Phys. {\bf A28}, 2005 (1995);
T. Sh. Misirpashaev, Sov. Phys. JETP {\bf 72}, 973 (1991);
M. Krusius, E.V. Thuneberg and U. Parts, Physica {\bf B197}, 376 (1994).

\bibitem{alexander} 
S. Alexander, R. Brandenberger, R. Easther and A. Sornborger,
hep-ph/9903254 (1999).

\bibitem{deLKraVac} A. A. de Laix, L. M. Krauss and T. Vachaspati,
Phys. Rev. Lett. {\bf 79}, 1968 (1997).

\bibitem{SmiVil} 
A. G. Smith and A. Vilenkin, Phys. Rev. {\bf D36}, 990 (1987).

\bibitem{Hin} 
G. Vincent, M. Hindmarsh and M. Sakellariadou,
Phys. Rev. {\bf D55}, 573 (1997).

\bibitem{Albetal} A. Albrecht, R. Battye and J. Robinson,
Phys. Rev. Lett., {\bf 79}, 4736 (1997);
Phys. Rev. {\bf D59}, 023508 (1998).

\bibitem{othercmbr}
D.P. Bennet, F.R. Bouchet and A. Stebbins, Nature {\bf 335} 410, (1988);
L. Perivolaropoulos, Phys. Lett. {\bf B298}, 305 (1993);
L. Perivolaropoulos, Ap. J. {\bf 451}, 429 (1995);
B. Allen, R. R. Caldwell, S. Dodelson, L. Knox,
E. P. S. Shellard and A. Stebbins, Phys. Rev. Lett. {\bf 79}, 2624 (1997);
U-L. Pen, U. Seljak and N. Turok, Phys. Rev. Lett. {\bf 79}, 1611 (1997);
P.P. Avelino, E.P.S. Shellard, J.H.P. Wu and B. Allen,
Phys. Rev. Lett. {\bf 81}, 2008 (1998);
R. Battye, J. Robinson and A. Albrecht,
Phys. Rev. Lett. {\bf 80}, 4847 (1998);
C. Contaldi, M. Hindmarsh and J. Magueijo, Phys. Rev. Lett. {\bf 82}, 
679 (1999);
E.J. Copeland, J. Magueijo and D.A. Steer, astro-ph/9903174.

\bibitem{peacock} ``Cosmological Physics'', J. A. Peacock,
Cambridge University Press (1999).


\end{thebibliography}
\end{document}